# Anomalous properties of the local dynamics in polymer glasses

R. Casalini[a,b] and C.M. Roland[a]

[a] Naval Research Laboratory, Chemistry Division, Code 6120, Washington DC 20375-5342, USA

[b] George Mason University, Department of Chemistry, Fairfax, VA 22030

The emergence of nanoscience has increased the importance of experiments able to probe the very local structure of materials, especially for disordered and heterogeneous systems. This is technologically important; for example, the nanoscale structure of glassy polymers has a direct correlation with their macroscopic physical properties. We have discovered how a local, high frequency dynamic process can be used to monitor and even predict macroscopic behavior in glassy polymers. Polyvinylethylenes vitrified by different chemical and thermodynamic pathways exhibit different densities in the glassy state. We find that the rate and amplitude of a high frequency relaxation mode (the Johari-Goldstein process involving local motion of segments of the chain backbone) can either correlate or anti-correlate with the density. This implies that neither the unoccupied (free) volume nor the configurational entropy governs the local dynamics in any general sense. Rather it is the magnitude of the *fluctuations in local density* that underlie these nm-scale motions. We show how properties of the dynamics and the density fluctuations can both be interpreted in terms of an asymmetric double

**well potential. Finally, since fluctuations underlie the macroscopic properties, we argue that information about the latter should be obtainable from characterization of the local dynamics.**

Polymers are fascinating molecules in which global (chain) and local (segmental) motions are convoluted, giving rise to a complex combination of behaviors and properties. As a consequence of its large size, a polymer molecule exhibits motions that span time scales ranging from nanoseconds for local segmental relaxation to years for macroscopic flow even in the liquid state. This means that a rubbery polymer exhibits many attributes of a "soft" solid, yet microscopically is indistinguishable from a liquid. The timescale of the segmental motions is extremely sensitive to proximity to the glass transition. On approaching the glass transition temperature, $T_g$, by cooling, the segmental motions slow by several orders of magnitude over a narrow temperature range, ultimately surpassing the duration of a typical experiment to become a "glass" (by definition). Various spectroscopic techniques are used to study the complicated dynamics of polymers, including mechanical measurements, nuclear magnetic resonance, light scattering, neutron scattering, and dielectric spectroscopy (DS). Among these DS has the important advantage of being able to cover more than 10 decades of time (or frequency), which is necessary to follow the multitude of relaxation processes of a polymer chain. In DS the segmental motion, comprised of correlated transitions of a couple of backbone bonds, is associated with structural (or the α–) relaxation, characterized by a relaxation time $\tau_\alpha$ that becomes extremely large in the glassy state. Its divergence below $T_g$ is still the subject of debate.[1,2]

In the glassy state other "secondary" processes (β, γ, …) are observed, having an origin commonly attributed to pendant groups undergoing local, uncorrelated reorientations. (These processes also transpire above $T_g$ but are difficult to resolve because of their small amplitude and overlap with the high frequency part of the α process). However, as pointed out by Johari and Goldstein[3], a secondary relaxation is observed for rigid molecules lacking side groups, which means there is a particular β relaxation, the Johari-Goldstein (JG) process, that is not simple side-group motion. The JG relaxation is a universal property of glass forming liquids and polymers, but nevertheless its molecular origin remains an open question. In at least one theoretical approach,[4] the JG is treated as the precursor of the structural α–relaxation. Alternatively, the JG relaxation is considered to be independent of the α–dynamics.[5] For example in energy landscape models, *"whereas the primary relaxation is assumed to be intrinsically coupled to transitions among different free-energy minima, this secondary relaxation process is viewed as a local relaxation within a given minimum"*.[6] Therefore the secondary relaxation is not necessarily related to the primary (structural) relaxation, although in energy landscape models possible correlations (e.g., between activation energy) could be supposed. An assumed independence of the primary and JG dynamics contrasts with recent DS measurements under hydrostatic pressure[7,8] and NMR results[9] demonstrating a strong correlation between the JG and α relaxations, notwithstanding their very different timescales. The relationship between the JG and segmental processes is analogous to that between the segmental and chain modes of polymers – in both cases there is a difference in length and time scales, even though the same molecular units are involved in both motions. Despite a growing recognition of the strong correlation between the α and the JG dynamics, there is no model that relates the behavior of the JG to other physical properties.

Herein we present a study of the JG relaxation in glassy polyvinylethylene (PVE), vitrified by different routes leading to different densities (three of these pathways are illustrated in Figure 1). PVE is a judicious choice of polymer for this study because of its chemical structure - the dipoles responsible for the JG relaxation are the same as those giving rise to the α relaxation. We find unexpected and somewhat unintuitive behavior; to wit, at a given temperature and pressure, the JG relaxation can be slower or faster for densities greater than a "standard polymer glass". That is, the JG dynamics can be correlated or anti-correlated with the available volume. We ascribe this behavior to the inherently heterogeneous character of the dynamics. Distinct length scales, with contributions reflected at different frequencies, respond differently and independently to volume changes. While properties such as structural relaxation in the liquid state are governed by the bulk volume, for the (very local) JG process the local distribution of the volume is paramount.

The PVE was 96% vinyl (1,2-addition product) polybutadiene obtained from Bridgestone Americas, with a weight-average molecular weight = $1.53 \times 10^5$ D. Dielectric measurements employed a high precision Andeen-Hagerling 2700A bridge. The instrument has an exceptionally high resolution (loss tangent = $1.5 \times 10^{-8}$) but a somewhat limited frequency range (50 Hz – 20 kHz). Measurements at atmospheric pressure were done in a custom built, closed-cycle helium cryostat (Cryo Ind.) with temperature stability better than $5 \times 10^{-2}$K over a week's duration. For elevated pressure measurements the sample and electrodes were in a high pressure cell inside an environmental chamber (Tenney Co.); the experimental configuration is described more fully elsewhere[8]. Network formation (a fourth route to density modification, not shown in Fig. 1) was accomplished by reaction of the PVE with 0.67% by weight dicumyl peroxide (Varox DCP-R from R.T. Vanderbilt); details of this procedure can be found elsewhere.[10] For all cases PVE films were measured between parallel plates, without spacers so that the plates could move freely in response to volume changes in the sample during the experiments. The JG relaxation time was calculated from the peak frequency ($\tau=(2\pi f_{max})^{-1}$) by fitting the dielectric loss spectra using a Cole-Cole relaxation function[11]; this $\tau_{JG}$ represents the most probable value.

The choice of PVE facilitates study of the JG relaxation, since its dielectrum spectrum shows an resolved, and thus unambiguous, JG secondary relaxation peak.[12,13] Fig. 1 illustrates three routes used to obtain glassy PVE, each resulting in a different density at the same $T$ and $P$. The first (sample A) consists of isobaric cooling at a constant rate (2K/min) from above $T_g$ to a temperature well below the PVE glass transition temperature ($T_g$ = 272.5 K[14]). Since the pressure is one atmosphere, this material is referred to as the "standard" PVE glass. The polymer is in a non-equilibrium state, having a volume in excess of the equilibrium value (indicated by the dotted line in Fig. 1). The dielectric loss spectrum for sample A is shown in Figure 2. Only the JG process is observed because structural relaxation is too slow ($\tau_\alpha > 10^4$s) to contribute in the available frequency window. As the sample is maintained under isobaric and isothermal conditions, the volume slowly relaxes toward thermodynamic equilibrium, yielding a progressively denser material (sample B); this process is called "physical aging". Dielectric loss spectra recorded at different aging times (Fig. 2) show a clear decrease in amplitude of the dispersion, *along with a shift to higher frequency*. The inset to Fig. 2 shows the change in the JG relaxation time versus aging time. It is remarkable that the increased density during aging (estimated to be about 0.04%) causes the JG process to *speed up*; that is, at constant $T$ $\tau_{JG}$ becomes *smaller* with decreasing volume. The amplitude of the JG peak also decreases during aging, as reported previously by several authors.[15-18] In a few prior studies of small molecules rather than polymers, the secondary relaxation time was found to be affected by physical aging well below $T_g$.[17,18,19,20,21] For these cases, either no appreciable change in $\tau_{JG}$ was observed, an expected consequence if the temperature is sufficiently below $T_g$ that physical aging is extremely slow,[15,18] or $\tau_{JG}$ was reported to decrease. We note that the decrease herein of the JG amplitude with density is also counterintuitive, since a denser material has a larger concentration of dipoles. And indeed, a larger value of the unrelaxed part ($f \gg \tau_{JG}^{-1}$) of the real part of the permittivity is observed. The decrease in the amplitude of the dielectric loss peak means that there is a decrease of the average reorientation.

Another route to glassy PVE makes use of hydrostatic pressure. It is known that cooling at high pressure to below $T_g$, followed by release to ambient pressure produces a denser polymeric glass.[22-24] In the present case the PVE is first pressurized to 350 MPa at 60K above $T_g$, with glass formation taking place during subsequent isobaric cooling. In the glassy state at constant temperature (242.8K), the pressure is then reduced to ambient. This material (sample C in Fig. 1), known as "pressure-densified" glass, has a larger density (almost 1% larger) than the glass formed conventionally by cooling at atmospheric pressure (sample A).

Comparing the dielectric loss spectra of samples A and C at atmospheric pressure and T=242.8K (Figure 3), the JG relaxation for the latter has a larger amplitude and occurs at lower frequency than the peak for sample A. Thus, pressure densification, which increases the density, causes the expected changes in the JG process – slowing down with concomitantly larger amplitude. Over the range of temperatures up through $T_g$, the JG relaxation time of sample C remains longer than $\tau_{JG}$ for A (Fig. 3 inset). The JG activation energy, $E_{JG}$, of C is slightly smaller than for A ($E_{JG}^A = 46 \pm 2\, kJmole^{-1}$; $E_{JG}^C = 42 \pm 2\, kJmole^{-1}$), while a substantial difference is observed in the high temperature limiting values of $\tau_{JG}$ ($\log(\tau_{JG}^A) = -14.1 \pm 0.3$; $\log(\tau_{JG}^C) = -12.8 \pm 0.2$). Note that physical aging during the course of these measurements is negligible. These changes in the JG process for PVE are consistent with previous observations for molecular liquids.[25,26]

Thermodynamic models would account for the slowing down of the JG relaxation in sample C in terms of a reduction in either the free volume or the configurational entropy. These are the accepted explanations of the equilibrium dynamic behavior[27,28] but they cannot explain the opposite effect that density has on sample B. The implication is that samples C and B are different, not only in terms of their macroscopic density, but also in their microscopic structure. For this reason a description of the dynamics in term of macroscopic thermodynamical quantities fails. Moreover, the sensitivity of the JG to the local structure is an indication that the conventional picture of the JG process in polymers as simple, non-cooperative relaxation of a single repeat unit is untenable. There must be a number of segments involved, dependent on the particular conditions under which the glass is formed. The fact that the local distribution of volume, rather than the macroscopic volume, governs the JG relaxation means that this process can be used to probe the local structure of the polymer, which is important for macroscopic properties, as we now describe.

Previous studies comparing physically aged and pressure-densified glassy polymers found interesting differences in their mechanical properties. In particular, physical aging confers brittleness while pressure densification increases the material's ductility.[29-31] In some cases the increased brittleness of aged glassy polymers can be removed ("rejuvenation") by mechanical stress.[31,32]

Interesting insight into the differences between glasses formed by different pathways is gleaned from X-ray scattering measurements on polystyrene.[33] Large scale density fluctuations were found to correlate with the macroscopic density, irrespective of the method used to vitrify the material; however, density fluctuations occurring over a smaller scale (< 0.5 nm) varied substantially depending on the pathway to the glass. Independent of the macroscopic density, physical aging increased local ordering. On the other hand, pressure densification had the opposite effect, with the reduction in local order correlated with improved mechanical properties, in particular producing a more ductile glass. Thus, there is an interesting connection between macroscopic and (sub-nm) microscopic properties.

If we assume that the motions giving rise to local density fluctuations in a glassy polymer are related to the same reorientational motions comprising the JG relaxation, certain behaviors can be anticipated. First, since the probability of a fluctuation is related to the minimum energy, $E_0$, needed to carry out reversibly a given change of density,[34] a system having larger fluctuations is characterized by a smaller value of $E_0$. If $E_0$ is related to the energy barrier that needs to be overcome to orient dipoles, then larger fluctuations correspond to a larger amplitude of the dielectric loss (and vice versa).

An interpretation of both the density fluctuations and their influence on the dielectric loss can be obtained using the asymmetric double well potential (ADWP)[5,35,36,37] model, used previously by Dyre and Olsen[5] to analyze JG relaxations in small molecule glass formers. The ADWP is described in terms of two quantities, the energy barrier $U$ and the asymmetry $\Delta$ (see insert fig.4). The relaxation time and the height of the dielectric loss peak, $\varepsilon''_{\max}$, are given by[5,35]

$$\tau_{JG} = \tau_0 \exp\left(\frac{2U+\Delta}{2k_B T}\right) \cosh^{-1}\left(\frac{\Delta}{2k_B T}\right)$$

$$\varepsilon''_{max} = \varepsilon''_0(T) \cosh^{-2}\left(\frac{\Delta}{2k_B T}\right) \quad (1)$$

The parameter $\tau_0$ is independent of both structure and temperature, while $\varepsilon''_0$ is independent of structure but varies inversely with temperature. The free energy differences $U$ and $\Delta$ are structure dependent and it is this dependence that can be related to the fictive temperature, $T_f$, characterizing the non-equilibrium glass[5]: As $T_f$ decreases, U decreases and $\Delta$ increases.

The ADWP model predicts at least qualitatively the behavior observed during aging. During isothermal aging $T_f$ decreases with aging time; therefore, $\Delta$ increases and $U$ decreases and consequently, according to eq.(1), $\varepsilon''_{max}$ and $\tau_{JG}$ both decrease (insert fig.2). As discussed by Dyre[5] the ADWP model gives a description of the quasi-linear relationship between $\log(\varepsilon''_{max})$ and $\log(\tau_{JG})$ shown in fig.4. From eq.(1) it is easy to see that the changes of $\varepsilon''_{max}$ and $\tau_{JG}$ between the aging times $t_1$ and $t_2$ are related as

$$\ln\left(\frac{\tau_{JG}^{t_1}}{\tau_{JG}^{t_2}}\right) = \frac{2(U^{t_1}-U^{t_2})+\Delta^{t_1}-\Delta^{t_2}}{2kT} + 2\ln\left(\frac{\varepsilon''_{max}{}^{t_1}}{\varepsilon''_{max}{}^{t_2}}\right) \quad (2)$$

from which it follow that when $2U+\Delta$ is constant during aging (which is roughly true since U and $\Delta$ have opposite behavior), the two quantities should exhibit a power relationship with an exponent expected to equal 0.5. As seen in fig.4, we obtain $d\log(\varepsilon''_{max})/d\log(\tau_{JG}) = {\sim}0.34$. The lower experimental value can be attributed to a decrease of $2U+\Delta$ during aging. The change of the potential energy barriers during aging can be calculated from eq.(1) as

$$\Delta^{t_1} - \Delta^{t_2} \cong -kT \ln\left(\varepsilon''_{max}{}^{t_1}/\varepsilon''_{max}{}^{t_2}\right)$$

$$U^{t_1} - U^{t_2} \cong kT\left[\ln\left(\tau_{JG}^{t_1}/\tau_{JG}^{t_2}\right) - \frac{3}{2}\ln\left(\varepsilon''_{max}{}^{t_1}/\varepsilon''_{max}{}^{t_2}\right)\right] \quad (3)$$

Using these equations we calculate that the changes in the spectra in fig.2 correspond to $U^B-U^A$= -0.32 kJ mole$^{-1}$ and $\Delta^B-\Delta^A$= 0.15 kJmole$^{-1}$. Thus, during aging the energy barrier decreases, the potential becomes more asymmetric, and 2U+Δ decreases.

The ADWP model can also be applied to the behavior observed during pressure densification. For sample C formed at higher pressure, since $T_g$ increases with pressure, its $T_f$ is higher than for the standard glass A. Consequently, Δ is smaller and U larger than for A, which means that $\varepsilon''_{max}$ and $\tau_{JG}$ are larger for sample C than for A. This is exactly the experimental results seen in fig.3. Using eqs.(1) we calculate for the changes in the spectra $U^C-U^A$=2.4 kJmole$^{-1}$ and $\Delta^C-\Delta^A$=-0.8 kJmole$^{-1}$. The pressure densified sample has a more symmetric potential with larger energy barrier.

It is reasonable to expect that local fluctuations are controlled by the same asymmetry of the potential , with the probability of a fluctuation increasing with decreasing Δ. For smaller Δ the difference between the transition rates between the two energy levels is smaller. In other words, the system becomes less ordered with greater probability of rearrangements. But since these rearrangements are local, they are not connected to the macroscopic density of the system.

A very different means to alter the density of a polymer is by chemically crosslinking the chains to form a network. Crosslinking was carried out on the PVE at atmospheric pressure well above $T_g$, with the material subsequently cooled isobarically to the glassy state. The formation of a network yields density increases of as much as 5% and also systematically increases the glass transition temperature (by about 15K for the highest crosslinking density).[38] As shown in Figure 5, the PVE network with the highest degree of crosslinking exhibits a JG relaxation significantly faster than for the linear PVE (sample A) but has a similar activation energy ($E_{JG}^{D} = 42 \pm 2\,kJmole^{-1}$, $\log_{10}(\tau_{JG}^{D}) = -13.9 \pm 0.3$ ). Thus, higher macroscopic density via crosslinking has the anomalous effect on the microscopic motion – denser PVE exhibits faster JG dynamics. (A comparison of the dielectric strength was not feasible because of experimental uncertainties.) Note also that since $T_g$ increases with crosslinking, making this comparison at the same $T$-$T_g$ would result in a even larger difference in the $\tau_{JG}$.

Since $E_{JG}$ remains almost unaltered, in the framework of the ADWP model the difference between the $\tau_{JG}$ of the linear and the crosslinked samples is due to either larger Δ or/and smaller U for the latter. An increase in Δ is attributable to the limited mobility of the segments in proximity of the crosslinkings. Although no X-ray data are available, it can be argued that network formation should hinder the small scale density fluctuations. Thus, there is again consistency with the experimentally observed correlation of density fluctuation and $\tau_{JG}$; i.e., smaller density fluctuation are correlated with smaller $\tau_{JG}$.

Like for the effect on the JG relaxation, crosslinking affects the mechanical properties of the PVE in similar fashion to that observed for physical aging, reduced extensibility leading to a brittle material. Measuring the JG relaxation for different degrees of crosslinking, we find that while the macroscopic density increases with crosslinking, as long as the material retains its property of high elasticity (distance between crosslinks > Kuhn length), $\tau_{JG}$ is unaffected. However, for higher degrees of crosslinking (crosslink distance < Kuhn length), $\tau_{JG}$ begins to decrease. Thus, again the properties of the microscopic JG motion relate directly to the macroscopic mechanical properties of the polymer. The anomaly of shorter $\tau_{JG}$ for larger density in the networks is associated with reduced ductility (extensibility), implying there is a concomitant decrease in the local density fluctuations.

The changes observed in the JG dynamics for different methods of vitrification provide interesting revelations about the nature of the JG process. It is clear that the JG properties are not governed by the macroscopic density (nor the entropy), but can be explained qualitatively in terms of change of the local energy potential using the ADWP model. We find that the JG motion follows the local (< 1 nm) density fluctuations and therefore can be used to probe them on the nanometer scale. This is important since, as we have seen, the nm scale density fluctuations have a direct bearing on macroscopic mechanical properties of glassy polymers, such as their ductility. The ADWP provides a qualitative understanding of the correlation between local density fluctuation and JG relaxation. A better understanding of this connection is technologically desirable, in view of the widespread use of polymers in their glassy state and the recent advances in manipulating materials at the nanometer level. Our results herein indicate that the JG relaxation can potentially track the local (nm) structure of the glass, so that the properties of the JG process can reveal structural and conformational information that is otherwise not easy to obtain. However, further studies and comparison with other techniques (e.g. X-ray) are necessary to gain a more quantitative understanding.

The authors acknowledge the support of the Office of Naval Research.

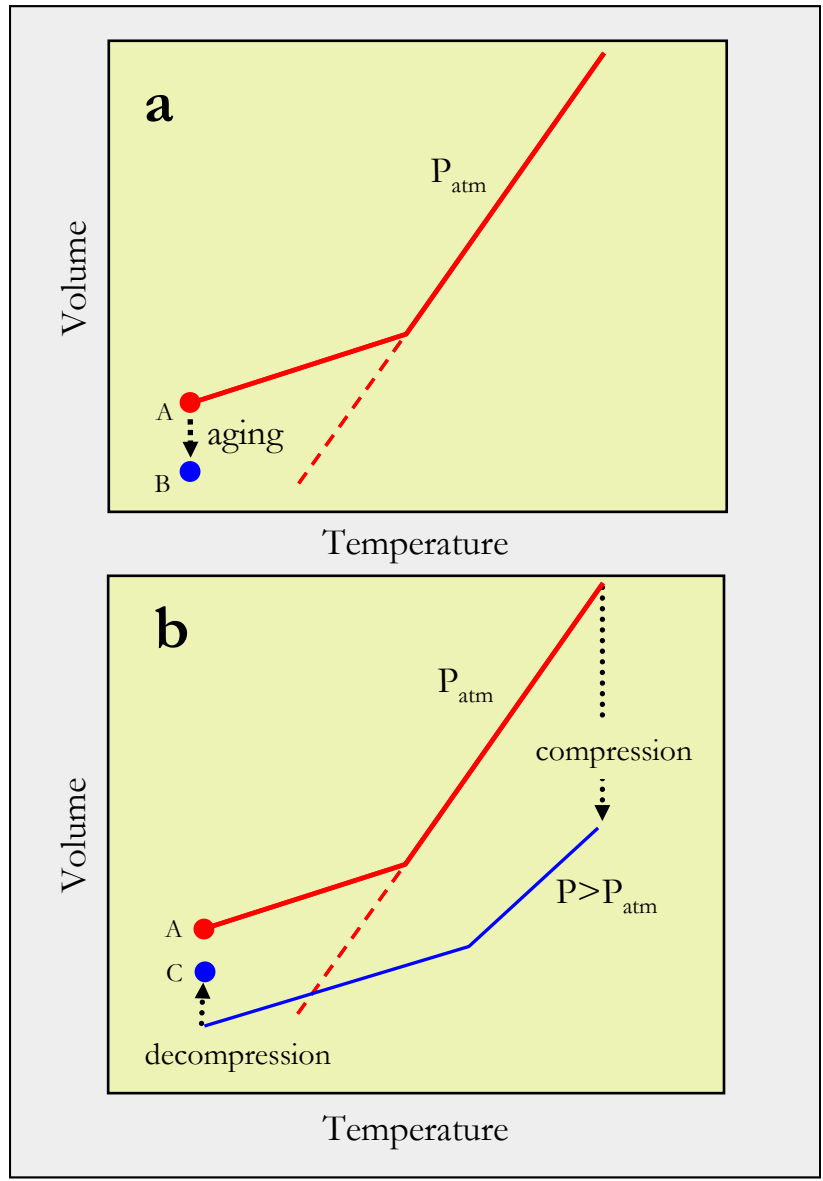

**Figure 1. Schematic of different routes to increase the density of a glassy polymer.** (a) Sample A ('standard glass') was cooled at a constant rate from the equilibrium state above the glass transition; sample B ('physically aged glass') was obtained by maintaining sample A at the same temperature for a time long enough for its volume to approach the equilibrium value (dotted line). (b) Sample A was formed as above, while sample C was obtained by cooling after a compression in the melt and subsequent decompression in the glassy regime ("pressure densified glass").

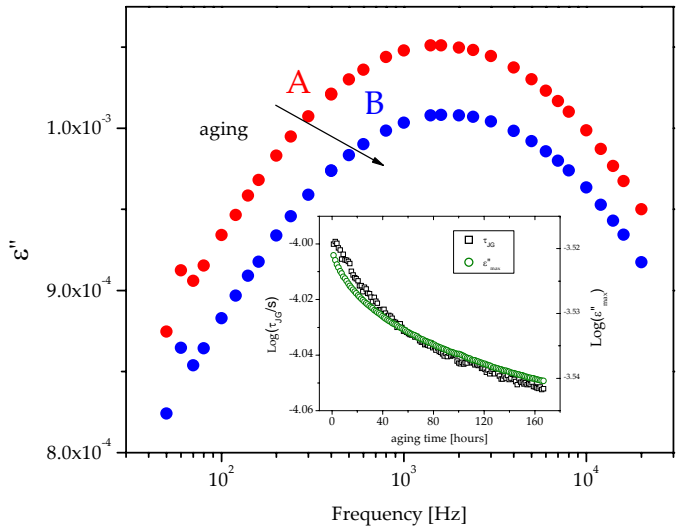

**Figure 2. Effect of physical aging on the JG relaxation (route a)**. Imaginary part of the permittivity for samples cooled below the glass transition at a constant rate and then maintained at T=240.9K. The difference between the curves is the time elapsed (163 hours) at 240.9K prior to the measurements. Insert: variation of the JG relaxation time, $\tau_{JG}$ and maximum of dielectric loss, $\varepsilon''_{max}$, during physical aging.

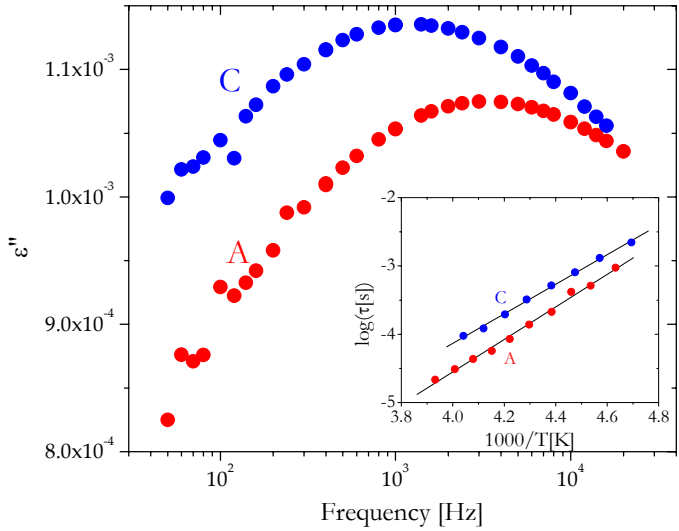

**Figure 3. Effect of high pressure on the JG relaxation (route b).** Imaginary part of the permittivity at atmospheric pressure and T=242.8K. Sample A was vitrified at atmospheric pressure (figure 1b), while C was pressure-densified (glass formation under high pressure) (figure 1b). Insert: comparison of the temperature dependence of the JG relaxation time.

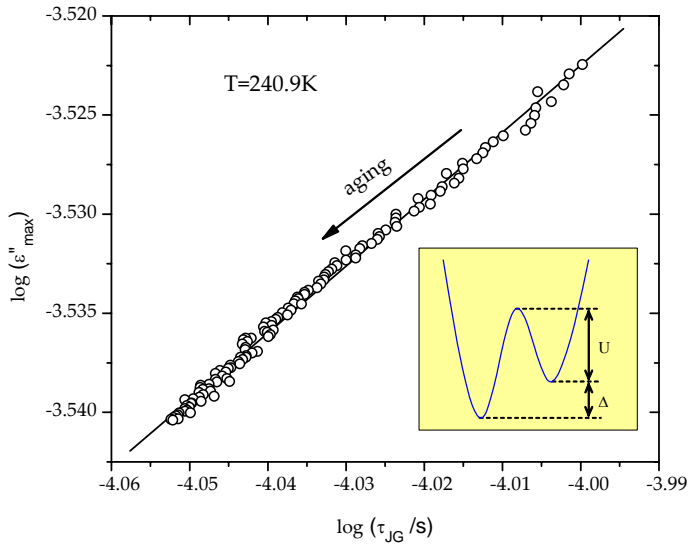

**Figure 4. Test of ADWP model.** Double logarithmic plot of the amplitude of the dielectric loss versus the JG relaxation time during isothermal aging. The solid line is a linear fit with a slope = 0.34. This power law behavior, $\varepsilon''_{max}\tau_{max}^{-0.34}$ = constant, is consistent with the results of Olsen et al.[19], interpreted using the ADWP (shown schematically in the inset).[5] From eq.(1) with the approximation that the sum $2U+\Delta$ does not change with aging (since U increases while $\Delta$ decreases), the slope should be 0.5. A linear coefficient smaller than 0.5 implies a decrease of $2U+\Delta$ with aging.

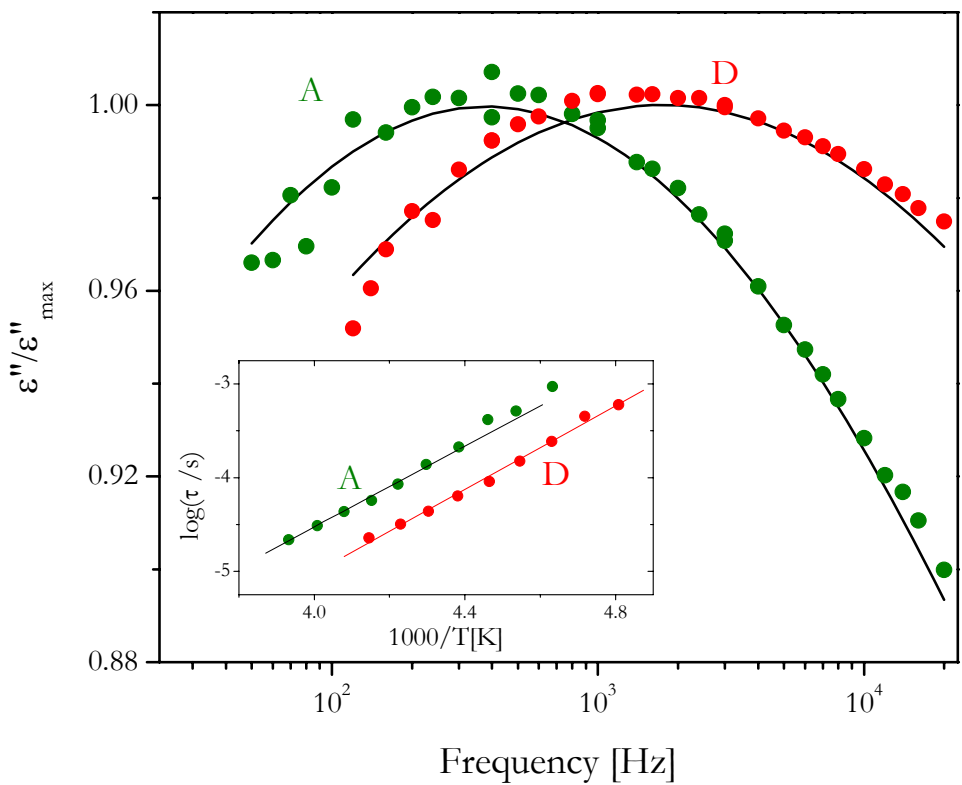

**Figure 4. Effect of chemical crosslinking on JG relaxation**. Imaginary part of the permittivity at atmospheric pressure and T=224K for the linear (A) and the crosslinked PVE (D). Both samples were cooled from the equilibrium state at the same rate (2K/min). Insert: comparison of the temperature dependence of the JG-process.